\documentstyle[aps,twocolumn]{revtex}
\begin{document}
\draft
\title{Gravitational radiation, energy and reaction 
on quasi-spherical black holes}
\author{Sean A. Hayward}
\address{Asia Pacific Center for Theoretical Physics,\\
The Korea Foundation for Advanced Study Building 7th Floor,
Yoksam-dong 678-39, Kangnam-gu, Seoul 135-081, Korea\\
and\\
Department of Physics, Konkuk University,
93-1 Mojin-dong, Kwangjin-gu, Seoul 143-701, Korea\\
{\tt hayward@mail.apctp.org}}
\date{Revised 5th February 2001}
\maketitle

\begin{abstract}
Gravitational radiation is locally defined 
where the wavefronts are roughly spherical.
A local energy tensor is defined for the gravitational radiation.
Including this energy tensor 
as a source in the truncated Einstein equations 
describes gravitational radiation reaction,
such as back-reaction on a roughly spherical black hole.
The energy-momentum in a canonical frame is covariantly conserved.
The strain to be measured by a distant detector is simply defined.
\end{abstract}

Gravitational-wave astronomy 
is expected to become a major observational science in the coming millenium.
Gravitational-wave theory, however, is based on various approximations,
as there is no general local definition of a gravitational wave
in General Relativity;
see e.g\ the review of Thorne\cite{T}.
Einstein\cite{E} initiated the study of gravitational waves,
in the linearized approximation on flat space-time,
but there were objections to their reality and ability to carry energy.
General belief came only in the 1960s with two further approximations.
Firstly, assuming asymptotic flatness allowed 
a definition of gravitational wave infinitely far from all sources,
as initiated by Bondi
and formalized in the conformal framework of Penrose\cite{Pe}.
Secondly, the high-frequency or short-wavelength approximation 
allowed a locally averaged definition of gravitational wave,
as formulated by Isaacson after initial ideas of Wheeler\cite{I,MTW}.
In both cases, the physical reality of the waves is indicated by 
definitions of their energy which, added to that of the matter,
yields energy-balance equations:
the Bondi energy-loss equation and the averaged Einstein equation, respectively.
This letter describes how this may also be achieved 
in a quasi-spherical approximation scheme.
Moreover, the resulting energy tensor of the radiation is local.
Thus the approximation allows 
a definition of gravitational radiation at any point.

The recent quasi-spherical approximation\cite{qs} 
was intended to be applicable to coalescing black-hole binaries,
one of the main expected sources for gravitational-wave detectors.
It may also prove applicable to neutron stars or supernovas,
or indeed any astrophysical situation which has rough spherical symmetry.
Further, 
for any other process which can be enclosed by roughly spherical surfaces,
it provides a mid-zone and, assuming isolation, far-zone approximation.
Mathematically this is achieved by linearizing certain fields 
which would vanish in exact spherical symmetry,
having made a decomposition of space-time 
adapted to the roughly spherical surfaces,
henceforth called transverse surfaces.
Since this makes no assumption of closeness to stationarity,
the approximation holds for arbitrarily fast dynamical processes.
The approximation has also been tested against angular momentum 
by applying it to Kerr black holes\cite{SH}:
the error in the strain waveform is much lower than 
expected signals from binary black-hole coalescence.

The wavefronts of outgoing and ingoing gravitational radiation 
form two families of null hypersurfaces,
intersecting in the two-parameter family of transverse spatial surfaces.
This geometry is described by the formalism of dual-null dynamics\cite{dn,dne},
briefly summarized here. 
Labelling the hypersurfaces by null coordinates $x^+$ and $x^-$, 
and taking coordinates $x^a$ for the transverse surfaces,
the space-time metric takes the form
\begin{eqnarray}
g&=&h_{ab}(dx^a+s_+^adx^++s_-^adx^-)\otimes\nonumber\\
&&(dx^b+s_+^bdx^++s_-^bdx^-)
-2e^{-f}dx^+\otimes dx^-
\end{eqnarray}
where $h$ is the metric of the transverse surfaces,
$s_\pm$ are two shift vectors and $f$ is a normalization function;
where appropriate, Latin letters are used for transverse indices
and Greek letters for space-time indices.
The covariant derivative of $h$ is denoted by $D$,
and the evolution derivatives are
\begin{equation}
\Delta_\pm=\bot L_{\partial/\partial x^\pm}
\end{equation}
where $L$ denotes the Lie derivative
and $\bot$ denotes projection by $h$.
It is also useful to decompose $h$ into 
a conformal factor $\Omega$ and a conformal metric $k$ by
\begin{equation}
h_{ab}=\Omega^{-2}k_{ab}
\end{equation}
such that the Hodge operator $\hat{*}$ of $k$ is given by
the standard spherical-polar area form
\begin{equation}
\hat{*}1=\sin\theta\,d\theta\wedge d\phi
\end{equation}
where $x^a=(\theta,\phi)$ are quasi-spherical polar coordinates.
Then $\Omega^{-1}$ is the quasi-spherical radius.
The extrinsic curvature of the dual-null foliation may be encoded in
conformally rescaled expansions $\vartheta_\pm$ 
and traceless shears $\varsigma_{\pm ab}$,
inaffinities $\nu_\pm$ and twist $\omega_a$, 
defined previously\cite{qs} and below as necessary.
The Einstein equation and contracted Bianchi identity 
may then be written in a first-order form expressing 
$\Delta_\pm$ derivatives of the dynamical fields 
$(\vartheta_\pm,\nu_\pm,\varsigma_\pm,\omega,\Omega,f,k,s_\pm)$ 
in terms of the dynamical fields 
and their first and second transverse derivatives.

The {\em first quasi-spherical approximation}, introduced previously\cite{qs},
consists of linearizing in $(\varsigma_\pm,\omega,s_\pm,D)$, 
which vanish in spherical symmetry\cite{sph,1st}.
This yields a greatly simplified set of truncated equations, 
decoupling into a three-level hierarchy.
Moreover, the last level, the equations for $(\omega,s_\pm)$, 
need not be solved for the radiation problem.
This is because, fixing $\Delta_+$ to be the outgoing derivative, 
the Bondi news at null infinity $\Im^\pm$ is 
essentially $\varsigma_\mp=\Omega^{-1}\Delta_\mp k$\cite{qs,SH,mono}.
(Generally, one should use the null derivatives $\Delta_\pm-\bot L_{s_\pm}$;
however, applied to transverse tensors,
they coincide with $\Delta_\pm$ in the quasi-spherical truncation,
so will not be distinguished in this article).
Specifically, the conformal strain tensor at future null infinity $\Im^+$ is
\begin{equation}
\varepsilon={1\over2}\int\varsigma_-\,dx^-.
\end{equation}
This means that
\begin{equation}
\epsilon={\varepsilon\over{r}}
\end{equation}
is the strain tensor at a large distance $r$ from the source,
so that the displacements to be measured by a gravitational-wave detector are
\begin{equation}
{\delta\ell\over\ell}=\epsilon_{ab}e^ae^b
\end{equation}
where the Cartesian basis vector $e$ is the direction of displacement.
Thus the variables of the quasi-spherical approximation 
are directly related to the observable strain;
no further far-zone approximation is required.

Moreover, 
the Bondi flux may be localized 
in terms of a gravitational-radiation energy flux,
derived below as the appropriate contraction of an effective 
{\em energy tensor of the gravitational radiation}:
\begin{equation}
\Theta_{\alpha\beta}={\langle\Delta_\alpha k,\Delta_\beta k\rangle
-{1\over2}g^{\gamma\delta}\langle\Delta_\gamma k,\Delta_\delta k\rangle 
g_{\alpha\beta}\over{32\pi}}
\end{equation}
where $\langle\alpha,\beta\rangle=k^{ab}k^{cd}\alpha_{ac}\beta_{bd}$,
$\Delta_a=0$ and units are such that Newton's gravitational constant is unity.
Explaining this is the main purpose of this letter.
Principally, 
$\Theta$ acts like a matter energy tensor in the truncated Einstein equations
and satisfies a covariant conservation law, conservation of energy.
Also, $\Theta$ satisfies 
the strong, dominant, weak and null energy conditions\cite{gwbh}.
Thus {\em gravitational radiation carries positive energy}.
It should be stressed that (i) mathematically, $\Theta$ is a genuine tensor,
but depends on the dual-null foliation, not just on the space-time;
(ii) the physical interpretation of $\Theta$ as energy requires 
the quasi-spherical approximation to be valid,
meaning that the transverse surfaces must indeed be roughly spherical.
This will not be made precise here,
as the range of validity of the approximation 
is not clear in advance and best explored in applications.
The intuitive meaning of roughly spherical should be clear by any standards.

In short, the quasi-spherical approximation allows a local definition of
the energy-momentum-stress of gravitational radiation,
and therefore of the radiation itself:
{\em gravitational radiation} is present at a given point 
if and only if $\Theta$ is non-zero there.
With the above orientation, 
there is outgoing radiation if and only if 
$\Delta_-k$ (equivalently $\varsigma_-$) is non-zero,
and ingoing radiation if and only if 
$\Delta_+k$ (equivalently $\varsigma_+$) is non-zero.
The terminology radiation instead of wave is generally preferable,
since $\Delta_\pm k$ need not be oscillatory.
Instead, frequency spectra may be defined by Fourier transformations of $k$ 
from $x^\pm$ (at constant $x^\mp$) to frequency $f_\pm$\cite{gwbh}.
Then a {\em gravitational wave} 
is gravitational radiation which is peaked in frequency space.
This reflects the different physical basis as compared to that of
the Isaacson high-frequency approximation\cite{I,MTW}.
In the quasi-spherical approximation, 
gravitational radiation is defined even if there is no typical frequency.

Writing the non-zero components
\begin{eqnarray}
&&\Theta_{\pm\pm}=||\Delta_\pm k||^2/32\pi\\
&&\Theta_{ab}=e^f\langle\Delta_+k,\Delta_-k\rangle h_{ab}/32\pi
\end{eqnarray}
where $||\alpha||^2=\langle\alpha,\alpha\rangle$,
the $\Theta_{\pm\pm}$ components are reminiscent of
the high-frequency approximation.
The $\Theta_{ab}$ components show that 
there is generally a transverse radiation pressure 
produced by a combination of ingoing and outgoing gravitational radiation,
as well as the expected radial radiation pressure.
The fact that $\Theta_{+-}=0$ may be interpreted as meaning that 
gravitational radiation is purely radiative and workless,
as for the massless Klein-Gordon field.

Apart from the fact that a gravitational wave generally has two polarizations,
as encoded in the two independent components of $k$,
$\Theta$ is analogous to 
the energy tensor of Einstein-Rosen gravitational waves\cite{cyl},
which takes the massless Klein-Gordon form 
in terms of a gravitational potential generalizing the Newtonian potential.
A complex gravitational-radiation potential may similarly be defined 
for $\Theta$, which then takes a scalar-field form\cite{gwbh}.

If $\Theta$ is to measure the energy of the gravitational radiation,
then {\em gravitational radiation reaction}, 
the back-reaction on the space-time,
should be described by including $\Theta$ as a matter energy tensor 
in the truncated Einstein equations.
In fact, there is a more logical way to formulate this,
suggested by noting that the gravitational radiation is encoded in 
the conformal shears $\varsigma_\pm$,
which are linearized in the first approximation.
Then retaining non-linear terms in $\varsigma_\pm$
should give a more accurate approximation 
for the gravitational-radiation sector of the theory.
Thus the {\em second quasi-spherical approximation,}
described in detail in a longer article\cite{gwbh},
consists of linearizing in $(\omega,s_\pm,D)$ only. 
The resulting equations also decouple, this time into only two levels,
with the last level for $(\omega,s_\pm)$ 
again being irrelevant to the radiation problem.
The remaining equations are, taking the vacuum case,  
\begin{eqnarray}
&&\Delta_\pm\Omega=-\textstyle{1\over2}\Omega^2\vartheta_\pm\label{eq1}\\
&&\Delta_\pm f=\nu_\pm\\
&&\Delta_\pm k=\Omega\varsigma_\pm\\
&&\Delta_\pm\vartheta_\pm=-\nu_\pm\vartheta_\pm
-\textstyle{1\over4}\Omega||\varsigma_\pm||^2\\
&&\Delta_\pm\vartheta_\mp
=-\Omega(\textstyle{1\over2}\vartheta_+\vartheta_-+e^{-f})\\
&&\Delta_\pm\nu_\mp
=-\Omega^2(\textstyle{1\over2}\vartheta_+\vartheta_-+e^{-f}
-\textstyle{1\over4}\langle\varsigma_+,\varsigma_-\rangle)\\
&&\Delta_\pm\varsigma_\mp
=\Omega(\varsigma_+\circ\varsigma_-
-\textstyle{1\over2}\vartheta_\mp\varsigma_\pm)\label{eql}
\end{eqnarray}
where $(\alpha\circ\beta)_{ab}=k^{cd}\alpha_{ac}\beta_{bd}$.
The first three equations effectively define 
$(\vartheta_\pm,\nu_\pm,\varsigma_\pm)$.
The first approximation may, of course, 
be recovered by linearizing in $\varsigma_\pm$.
The additional quadratic terms in $\varsigma_\pm$ appear in the same way that 
a matter energy tensor with the form of $\Theta$ would,
as can be seen by comparing with 
the spherically symmetric equations\cite{sph,1st}.
More formally, one may introduce a truncated Einstein tensor $C$\cite{gwbh} 
in the first approximation.
Then the truncated Einstein equations are $C=8\pi T$ 
in the first approximation, where $T$ is the energy tensor of the matter,
and $C=8\pi(T+\Theta)$ in the second approximation.
This is the first reason for 
identifying $\Theta$ as an energy tensor for the gravitational radiation:
it plays the role of an effective matter energy tensor
in the second approximation.

Mathematically, the difference between first and second approximations is that 
in the first approximation,
the equations for $(\vartheta_\pm,\nu_\pm,\Omega,f)$,
the variables which survive in spherical symmetry,
decouple from the equations for $(\varsigma_\pm,k)$,
which constitute a wave equation for $k$. 
Physically this describes gravitational-wave propagation 
on a quasi-spherical background.
The background is not fixed in advance
and need not be spherically symmetric,
so even the first approximation is widely applicable.
There is no such decoupling in the second approximation:
the gravitational-radiation terms $(\varsigma_\pm,k)$ 
now enter the equations for the quasi-spherical part of the geometry,
which thus reacts to the passage of the waves.
Gravitational radiation reaction has thereby been included;
there is no longer a background which is independent of the waves.

Nevertheless, 
both first and second approximations share the remarkable feature that,
to compute the observable waveforms,
no transverse derivatives need be considered.
The truncated equations form an effectively two-dimensional system,
to be integrated independently at each angle of the sphere.
Physically this means that the observed gravitational-wave signal 
depends only on the line of sight to the source, surely a plausible result.
Moreover, by virtue of the dual-null formulation,
the equations are already written in characteristic form,
the mathematically standard form for analysis of hyperbolic equations.
Numerical implementation is consequently straightforward 
and computationally inexpensive.
Numerical codes exist for both first and second approximations\cite{SH}.

The second reason for identifying $\Theta$ as an energy tensor is that, 
added to the energy tensor of the matter, 
it yields a covariant energy conservation law.
This is derived in general in the longer article\cite{gwbh}
and in the vacuum case as follows, by a more direct method.
As in spherical symmetry\cite{sph,1st},
there is a canonical flow of time defined by the vector or 1-form
\begin{equation}
\xi={*}d\Omega^{-1}
\end{equation}
where $*$ is the Hodge operator of the evolution space, 
${*}1=e^{-f}dx^+\wedge dx^-$.
Its non-zero components are
\begin{equation}
\xi_\pm=\mp\Delta_\pm\Omega^{-1}.
\end{equation}
Then $\xi$ is analogous to 
the Killing vector of a stationary space-time.
The {\em energy-momentum density of the gravitational radiation},
referred to the canonical flow, 
is the vector or 1-form
\begin{equation}
j_\alpha=-\Theta_{\alpha\beta}\xi^\beta
\end{equation}
with non-zero components
\begin{equation}
j_\pm=\pm e^f\Theta_{\pm\pm}\Delta_\mp\Omega^{-1}.
\end{equation}
This has the same form as the energy-momentum density of the matter 
in spherical symmetry\cite{sph,1st}.
The corresponding {\em energy flux of the gravitational radiation} 
is the dual 1-form
\begin{equation}
\psi={*}j.
\end{equation}
The conformally rescaled flux 
\begin{equation}
\varphi=\Omega^{-2}\psi
\end{equation}
then has non-zero components 
\begin{equation}
\varphi_\pm=-{e^f\vartheta_\mp||\varsigma_\pm||^2\over{64\pi}}.
\end{equation}
These expressions have the same form as 
those for the Bondi flux at $\Im^\mp$\cite{qs,SH,mono}.
Thus the conformal flux $\varphi$ 
is a local generalization of the Bondi flux.
Denoting the Hodge operator of $g$ by $\star$, 
the quasi-spherical truncation identity
\begin{equation}
\star1={*}1\wedge\hat{*}\Omega^{-2}
\end{equation}
allows the divergence of $j$ to be written as
\begin{equation}
\nabla_\alpha j^\alpha={\star}d{\star}j=\Omega^2{*}d(\Omega^{-2}{*}j)
=\Omega^2{*}d\varphi.
\end{equation}
The truncated Einstein equations (\ref{eq1}--\ref{eql}) yield
\begin{equation}
\Delta_\pm\varphi_\mp={e^f\Omega\over{64\pi}}
\left(\vartheta_+\vartheta_-\langle\varsigma_+,\varsigma_-\rangle
+\textstyle{1\over4}||\varsigma_+||^2||\varsigma_-||^2\right)
\end{equation}
and therefore
\begin{equation}
{*}d\varphi=e^f(\Delta_-\varphi_+-\Delta_+\varphi_-)=0.
\end{equation}
Thus $j$ is covariantly conserved:
\begin{equation}
\nabla_\alpha j^\alpha=0.
\end{equation}
Physically this represents {\em conservation of energy},
as in spherical symmetry\cite{sph,1st}.
In the presence of matter, it holds for the combined energy 
of the gravitational radiation and matter\cite{gwbh},
as in cylindrical symmetry\cite{cyl}.
The Noether charge associated with the Noether current $j$ 
provides a definition of active gravitational mass-energy 
generalizing that of spherical symmetry,
including the energy of the gravitational radiation.
Then energy conservation can be written in the form of a first law\cite{gwbh}.

In summary, the quasi-spherical approximation scheme allows
local definitions of gravitational radiation and its energy.
The radiation is superimposed on a background in the first approximation,
but reacts back on the space-time in the second approximation.
Comparing results of the first and second approximations,
in a given situation,
provides a valuable internal guide to their accuracy,
independent of any other estimates.
In particular, 
the results obtained for Kerr black holes are numerically indistinguishable 
in the first and second approximations\cite{SH}.

Gravitational-wave theory has progressed since the review of Thorne\cite{T} 
by higher-order post-Newtonian approximations\cite{Bl}
and close-limit approximations\cite{Pu},
which can be used to describe, respectively, the pre-coalescence 
and post-coalescence phases of a binary black-hole inspiral.
The coalescence phase 
has long been thought to be tractable only by numerical methods,
but despite great efforts\cite{BBH}, the desired waveforms are not yet known.
The quasi-spherical approximation is intended to apply to 
the post-coalescence phase, thereby reducing the period for which 
full numerical simulations are required.
Similarly, a quasi-equilibrium approximation has also recently been suggested 
for the later pre-coalescence phase\cite{DMS}.

Gravitational radiation reaction is an essential physical ingredient 
in the post-Newtonian and quasi-equilibrium approximations,
but is not included in the close-limit approximation,
or indeed any other perturbative approach with a fixed background. 
In comparison, the second quasi-spherical approximation allows 
a fully relativistic inclusion of radiation reaction 
for rapidly evolving black holes.
It is applicable from the very moment of coalescence,
meaning the appearance of trapped surfaces 
enclosing both original trapped regions.

As an illustration, 
the following scenario is plausible on physical grounds:
a distorted black hole generally emits gravitational radiation, 
which generally backscatters to produce ingoing radiation,
which is absorbed by the black hole,
thereby increasing its mass and area and generally changing its shape,
thereby also absorbing some outgoing radiation,
thereby changing the profile of subsequent gravitational-radiation emission,
and so on by feedback.
This entire process is actually described, step by step, by
the truncated Einstein equations (\ref{eq1}--\ref{eql}).
The area-increase property was derived as 
the second law of black-hole dynamics\cite{bhd},
where a black hole is locally defined by a type of trapping horizon,
which was also shown to be achronal and have topologically spherical sections.
Corresponding spherically symmetric first\cite{1st} 
and zeroth\cite{mg9} laws of black-hole dynamics 
admit quasi-spherical generalizations\cite{gwbh}
involving a local definition of surface gravity and the above mass.

Thus there is an astrophysically realistic context in which 
both black holes and gravitational radiation are locally defined, 
along with their physical attributes,
with each influencing the other.
Apart from practical applications to gravitational-wave astronomy, 
this provides a rich arena 
in which to advance theoretical understanding of 
the dynamical interaction between gravitational radiation and black holes.

\bigskip\noindent
Acknowledgements.
Thanks to Hisa-aki Shinkai and Jong Hyuk Yoon for discussions.
Research partly supported by the APCTP visiting program
and a research grant of Konkuk University.

\end{document}